\documentclass[pra,aps,twocolumn,superscriptaddress]{revtex4}

\usepackage{graphicx}% Include figure files

\usepackage{bm}% bold math
\usepackage{amsmath}
\usepackage{amsfonts}
\usepackage{amssymb}
\usepackage{color}
\usepackage{units}

\newcommand{\bra}[1]{\left<#1\right|}
\newcommand{\ket}[1]{\left|#1\right>}
\newcommand{\up}{\uparrow}
\newcommand{\down}{\downarrow}

\begin{document}
\title{Dissipative quantum control of a spin chain}
\author{Giovanna Morigi}
\affiliation{Theoretische Physik, Universit\"at des Saarlandes, D-66123 
Saarbr\"ucken, Germany}

\author{J\"urgen Eschner}
\affiliation{Experimentalphysik, Universit\"at des Saarlandes, D-66123 
Saarbr\"ucken, Germany}

%--------------------------------------
\author{Cecilia Cormick}
\affiliation{IFEG, CONICET and Universidad Nacional de C\'ordoba, Ciudad
Universitaria,
X5016LAE, C\'ordoba, Argentina}

%--------------------------------------
\author{Yiheng Lin}
\affiliation{National Institute of Standards and Technology,
325 Broadway, Boulder, CO 80305, USA}

\author{Dietrich Leibfried}
\affiliation{National Institute of Standards and Technology,
325 Broadway, Boulder, CO 80305, USA}

\author{David J. Wineland}
\affiliation{National Institute of Standards and Technology,
325 Broadway, Boulder, CO 80305, USA}

\date{\today}
 
\begin{abstract}
A protocol is discussed for preparing a spin chain in a generic many-body state in the asymptotic limit of tailored non-unitary dynamics. The dynamics require the spectral resolution of the target state, optimized coherent pulses, engineered dissipation, and feedback. As an example, we discuss the preparation of an entangled antiferromagnetic state, and argue that the procedure can be applied to chains of trapped ions or Rydberg atoms. 
\end{abstract}
\maketitle

The robust generation of entangled states is a cornerstone in quantum technological applications. Multipartite entanglement, in fact, plays a crucial role in various tasks of quantum information and communication \cite{Horodecki,Hillery,Josza,Bruss}, and can enable one to achieve unprecedented levels of precision in sensor systems \cite{Qsensor} and metrology \cite{Qmetrology}. High-fidelity generation of multi-partite entangled states has been demonstrated with protocols based on deterministic operations \cite{Ladt}. These protocols realize dynamics which optimally couple an initial state to the target state with a sequence of unitary operations. Their performance becomes more challenging as the number of components $N$ increases. In fact, this often implies a larger number of high-fidelity operations, which makes the protocol more sensitive to parameter fluctuations and to disorder, and requires longer time scales, over which the detrimental effects of intrinsic noise and decoherence become more relevant. This situation has motivated the search for alternative strategies. 

One promising approach is based on engineering noise \cite{DallaTorre} and dissipation \cite{Poyatos,Diehl,Kraus,Verstraete} in order to drive a many-body system towards the desired non-classical target state. %This is achieved by means of tailored non-unitary dynamics of which the target state is the stationary state \cite{DallaTorre,Poyatos,Diehl,Kraus,Verstraete}. 
Protocols based on this idea are often denoted by quantum reservoir engineering (QRE), and their hallmark is the robustness against parameter fluctuations, which results from the non-unitary nature of the processes that pump the system into the target state. When based on dissipation, they can be considered a many-body generalization of optical pumping, originally proposed by Kastler for creating spin polarized atomic ensembles by means of spontaneous decay \cite{Kastler}. As in optical pumping, the target state is stable, effectively decoupled from the mechanism which pumps out all other states involved in the dynamics, but fed by the dissipative processes \cite{Davidovich,Kim}. Under these premises the population of the target state will increase asymptotically towards unity.  

The formal procedure for implementing QRE is usually based on constructing a Liouvillean $\mathcal L$ for the density matrix $\rho$ of the system, for which the target state $\varrho_T$ is the unique stationary state, i.e. $\mathcal L\varrho_T=0$ \cite{Kraus,Viola}. When $\varrho_T=\ket{\psi_T}\bra{\psi_T}$, then the condition can often be cast in terms of rate equations, which couple the population of the target state, $P_T={\rm Tr}\{\varrho_T\rho(t)\}$, with the populations $P_n$ of the states $|\psi_n\rangle$, forming together with $|\psi_T\rangle$ a complete and orthogonal basis in the state space. Denoting by $\Gamma_{n\to m}>0$ the rate coefficients for the transitions $|\psi_n\rangle\to|\psi_m\rangle$, the equation for $P_T$ reads
\begin{equation}
\label{Rate}
\dot{P}_T=-\Gamma_T P_T+\sum_{n\neq T} \Gamma_{n\to T}P_n\,,
\end{equation}
and the loss rate of  state $|\psi_n\rangle$ is $\Gamma_n=\sum_m\Gamma_{n\to m}$. The objective is to achieve $P_T\to 1$ as $t\to\infty$. From considerations based on detailed balance, it can be verified that a necessary condition for efficient production of $\varrho_T$ is $\Gamma_n\gg\Gamma_T$ for $n\neq T$. It is sufficient when $\min_{n\neq T,m}(\Gamma_{n\to m})\gg \Gamma_T$, which can be reached by exploiting symmetries of the dynamics \cite{Lidar}.  This idea is at the basis of several proposals for dissipatively pumping spin or harmonic-oscillator systems into bi-partite and into specific multi-partite entangled states; examples are found in Refs.~\cite{Davidovich, Davidovich:01, PlenioHuelga, Pielawa:07, Kim, Kastoryano, Molmer:13, CarrSaffman:13, Cormick:13, Moelmer:14, Carvalho:14}. Experimental demonstrations include realizations with trapped ions \cite{Blatt, Lin,Home}, atomic ensembles \cite{Krauter}, and superconducting qubits \cite{Devoret:2013}. The identification of the procedure, however, becomes more complex for arbitrary multi-partite entangled target states.  This calls for the development of viable protocols for the dissipative preparation of a generic entangled state of a many-body system. 

Here we discuss such a procedure. Our idea is to tailor the excitation spectrum of the many-body system, such that the target state is an eigenstate and all transitions between pairs of states can ideally be individually addressed. Engineered dissipation allows one to perform irreversible population transfer, in order to construct dynamics as in Eq.~\eqref{Rate}. Transitions from the target state are far off resonance from all pumping processes, so that the outcoupling rate $\Gamma_T$ is sufficiently small. By repeatedly applying a sequence of pulses that empty all other states, the system is pumped into the target state, with an asymptotic fidelity that depends on the ability to tailor the transition rates. This procedure generalizes a method for quantum-state preparation of molecules \cite{Morigi07} to many-body systems. It provides a complementary approach to the one proposed in Ref. \cite{Reiter:2015}. In addition, in order to counteract noise and decoherence, which become more and more important as the number of components increases, measurements followed by feedback operations are built into the pulse sequence which restore the effectiveness of the procedure for long evolution times. 

We illustrate the procedure by discussing the preparation of a spin chain in the entangled antiferromagnetic state,
\begin{equation}
\label{Target}
|\psi_T\rangle=\left(|\up\down\up\down\ldots\rangle+|\down\up\down\up\ldots\rangle\right)/\sqrt{2}\,,
\end{equation}
where $|\up\rangle$ and $|\down\rangle$ are the two energy eigenstates of a (pseudo) spin $1/2$, separated by $\hbar\omega_0$. For $N=2$ ions, $|\psi_T\rangle$ is a triplet (Dicke) state, which can be perfectly decoupled from collective spin excitations via quantum interference processes  \cite{Dicke}. One can thus construct dynamics for which $|\psi_T\rangle$ (or the corresponding singlet state) is stationary; examples are in Refs.~\cite{Beige, Kastoryano, CarrSaffman:13, Schaetz}. For $N>2$, however, this procedure cannot be directly applied, since $|\psi_T\rangle$ is no longer a Dicke state. 

In order to realize non-unitary dynamics of which $\ket{\psi_T}$ is the stationary state, one could construct a harmonic Hamiltonian of which $|\psi_T\rangle$ is the ground state. Dissipative preparation into the target state would then proceed by means of a generalization of sideband cooling \cite{Sideband}. Implementing these dynamics with spins would in general require one to work with an equidistant energy spectrum whose excitations are collective spin states, thus in general Dicke states. It has the drawback that, as $N$ grows, the number of undesired "dark states" that are decoupled via quantum interference increases and that disorder and inhomogeneities may render the spectrum anharmonic. 

\begin{figure}
\includegraphics[width=8.5cm]{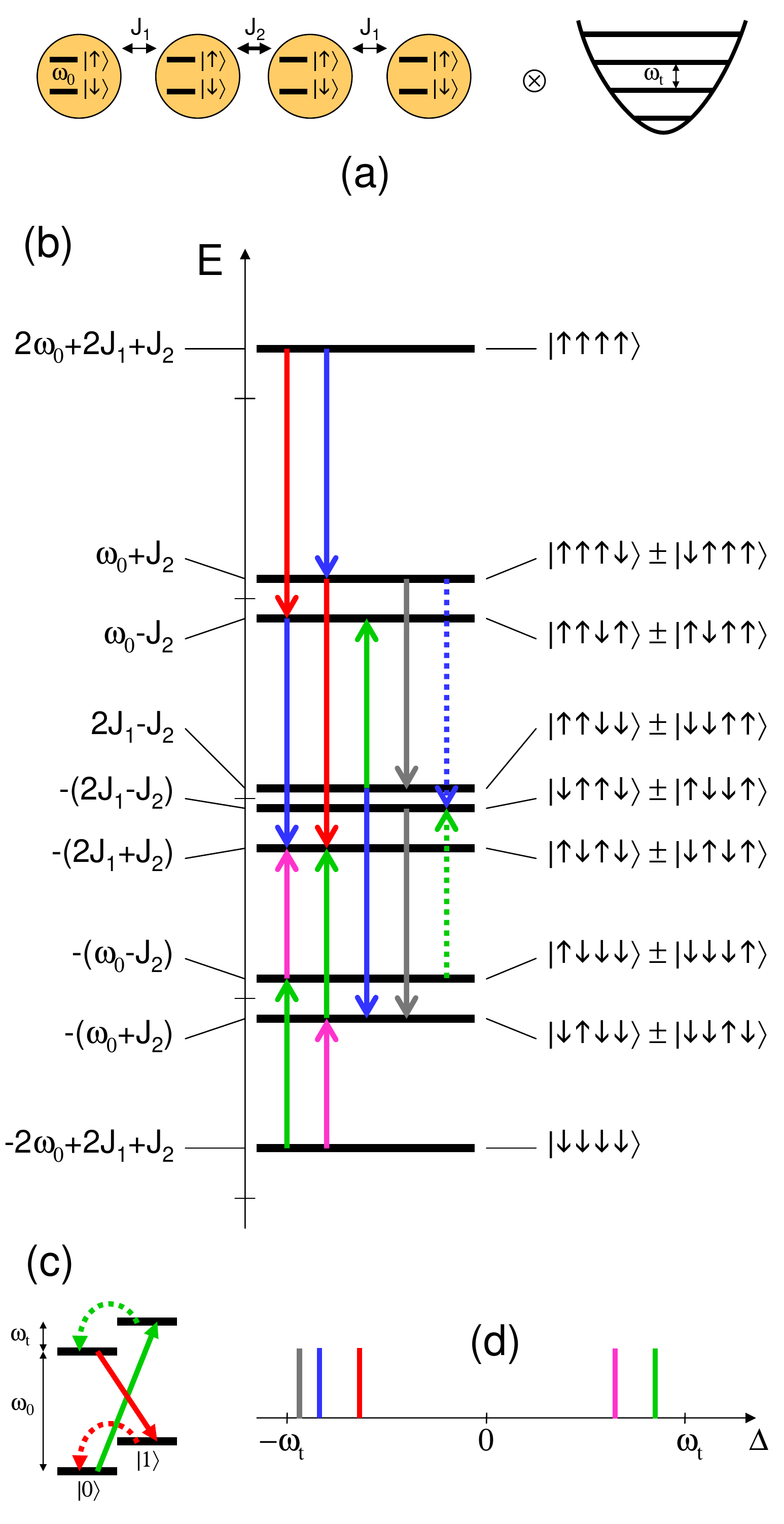}
\caption{Procedure for preparing the spins of a four-ion chain in the entangled state $(|\up\down\up\down\rangle+|\down\up\down\up\rangle)/\sqrt{2}$ via non-unitary dynamics. (a) The spins interact with one another ($J_j$) and with an ancilla, here a harmonic oscillator of frequency $\omega_t$. (b) The energy spectrum of the spin chain is tailored by controlling the couplings $J_j$ (only the $|n=0\rangle$ manifold of the ancilla is shown). Arrows indicate the pulses which resonantly pump the spins into the target state; each arrow represents two operations, as illustrated in (c): one coherent pulse (solid) that entangles spins and ancilla, and one dissipative pulse (dashed) that makes the transfer from state to state irreversible. In (d) the pulse frequencies corresponding to the arrows in (b) are displayed, using the same colour code. $\Delta$ denotes the detuning from $\omega_0$. The dashed arrows in (b) indicate transitions driven by the "blue" and "green" pulses that resonantly pump the spins into a state that is not the target. This state is depopulated by the "grey" pulse. %See Fig.~\ref{Fig:2} for more details on the pulse sequence.
}
\label{Fig:1}
\end{figure}

Instead, we choose a Hamiltonian whose spectrum is purposely tailored to be anharmonic and of which $|\psi_T\rangle$ is an eigenstate, but not necessarily the ground state. The interaction Hamiltonian
\begin{equation}
\label{H:int}
H_{\rm int}^{(N)}=\sum_{j=1}^{N-1} J_j\sigma_j^z\sigma_{j+1}^z
\end{equation}
serves this purpose, with $J_j>0$ and $\sigma_j^z$ the Pauli operator for spin $j$. This Hamiltonian can be realized for chains of Rydberg atoms \cite{Malossi:2014} or of trapped ions \cite{Deng}. In the latter case it is implemented by tailoring the coupling between the internal and the external degrees of freedom of the chain: the coupling coefficients $J_j$ depend on the interparticle distances and are symmetric about the center: in a linear Paul trap $J_{N-i}=J_i$ and $J_i<J_{i+1}$ for $i=1,\dots,[\frac{N}{2}]$. Fig.~\ref{Fig:1}(a) shows the specific case $N=4$.

Hamiltonian \eqref{H:int} stabilizes the target state and identifies the states $\psi_n$ entering the rate equations, whose coefficients $\Gamma_{n \to m}$ shall be engineered. The transitions $|\psi_n\rangle \to |\psi_m\rangle$ are driven resonantly by laser pulses, whereby the detunings $\Delta$ from the spin transitions vary from pulse to pulse. The corresponding spin Hamiltonian in the rotating frame reads
\begin{equation}H_0(\Delta)=-\hbar\Delta\sum_j\sigma_j^z/2\,,\end{equation}
of which $\ket{\psi_T}$ is eigenstate for $N$ even \cite{Footnote}. The resonant transitions for $N=4$ are shown in Fig.~\ref{Fig:1}.

The desired asymmetry in the coefficients $\Gamma_{n\to m}$ and $\Gamma_{m\to n}$  is achieved by means of engineered dissipation, along the lines of Ref.~\cite{Kastoryano}. It is realised by first entangling a single spin with an ancilla, followed by dissipation on the ancilla (see Fig.~\ref{Fig:1}(a,c)). In our example the ancilla is a harmonic oscillator, which may be a lossy cavity mode coupling with the spins 
%by means of appropriately tailored Jaynes-Cummings dynamics 
\cite{Kastoryano}, or a collective vibrational mode of an ion chain, coupling with the spins via the mechanical effects of light and sympathetically cooled to the ground state as in Ref.~\cite{Lin}.  Denoting by $|0\rangle$ and $|1\rangle$ the oscillator's ground and first excited state, irreversible population transfer from $|\psi_n\rangle$ to $|\psi_m\rangle$ is performed by first coherently driving the transition $|\psi_n\rangle|0\rangle\to |\psi_m\rangle|1\rangle$, and then damping the mode by an external reservoir that induces the transition $|\psi_m\rangle|1\rangle\to |\psi_m\rangle|0\rangle$. 

The coherent laser-driven dynamics which entangle spins and ancilla are described by the Hamiltonian %\cite{Cormick:13}
\begin{equation}
\label{H:L}
H(\Delta)=H_{\rm int}^{(N)}+H_0(\Delta)+\hbar  \sum_j g_j \sigma_j^x(a+a^\dagger)+\hbar\omega_t a^\dagger a\,.
\end{equation} 
Here, $g_j$ is the Rabi frequency, whose value is sufficiently small in order to drive only resonant transitions, and $a$ is the annihilation operator of the harmonic oscillator at frequency $\omega_t$.  %For simplicity we take $g_j=g$, but this is not a necessary condition. 
The oscillator is cooled at rate $\gamma$ to a steady-state excitation number $\bar n\ll 1$; the non-unitary cooling dynamics are described by the superoperator
 \cite{Gardiner} 
\begin{equation}
\label{Lgamma}
{\mathcal L}_{\gamma}\rho=\gamma(\bar n+1)\mathcal D[a]\rho+\gamma\bar
n\mathcal D[a^\dagger]\rho\,,
\end{equation}
where $\mathcal D[X]$ is a functional of the operator $X$ such that $\mathcal D[X]\rho=X\rho X^\dagger-(X^\dagger X\rho+\rho X^\dagger X)/2$, and $\rho$ is the density matrix of spins and ancilla. Pumping into the target state is realized by sequences of pulses, whose components correspond to the map $\mathcal T(\Delta,t)={\rm e}^{\mathcal L_{\gamma} t_{\gamma}}{\rm e}^{\mathcal L_{\rm coh}(\Delta) t}$, which alternates Liouvillian 
$\mathcal L_{\rm coh}(\Delta)\rho=[H(\Delta),\rho]/(i\hbar)$ for time $t$  with engineered dissipation as in Eq.~\eqref{Lgamma} for time $t_{\gamma}$.  The protocol iterates the concatenated map 
\begin{equation}
\mathcal T=\mathcal T(\Delta_j,
t_j)\mathcal T(\Delta_{j-1},t_{j-1})\ldots\mathcal T(\Delta_1,t_1)\,,
\end{equation}
where the sequence, the detunings and the durations $t_j$ are optimized to drive the system into the desired asymptotic state with close-to-unit fidelity. Over the time scale of a sequence, the dynamics can be cast in terms of a rate equation as in Eq.~\eqref{Rate} and the choice of the detunings results in tailoring the effective coefficients.  The idea might be regarded as a dissipative extension of the Law-Eberly protocol \cite{LawEberly}, originally developed for arbitrary quantum state preparation by means of coherent dynamics, and based on identifying the individual steps which deterministically connect an initial and a final state. Indeed, with our procedure, we achieve $\varrho_T=\lim_{\ell\to\infty}{\rm Tr}_{\rm an}\{{\mathcal T}^\ell\rho(0)\}$ for a certain set of initial states $\rho(0)$, where ${\rm Tr}_{\rm an}$ denotes the trace over the ancilla's degrees of freedom. 
Nevertheless, we find that the pumping efficiency considerably drops if the initial state is an equal statistical mixture and/or if the Rabi frequencies $g_j$ vary significantly. In fact, in these cases pumping happens into both $\ket{\psi_T}$ and the degenerate antisymmetric superposition $\left(|\up\down\up\down\ldots\rangle-|\down\up\down\up\ldots\rangle\right)/\sqrt{2}$. 
One remedy could be to alternate Hamiltonian $H(\Delta)$ with another pumping Hamiltonian, assuming one %has the capability to 
can engineer the spatial gradient of the pulse phase, but this approach is not robust against decoherence and fluctuations in the 
values of the couplings $g_j$. 

Our solution that enables pumping from arbitrary initial states into the target state is to include in the sequence a parity-correcting operation based on the protocol of Refs.~\cite{SorensenMolmer:99,Leibfried:04}. It performs a parity measurement, described by the operator 
\begin{equation}
\Pi=\sigma_1^x\ldots \sigma_N^x\,,
\end{equation} 
followed by a conditional operation on the system which corrects the parity in the case that it is not the desired one.  The corresponding dynamics can be realized by means of an ancilla, whose relevant states are denoted by $|A\rangle$ and $|B\rangle$: given the state of the system is $|\psi\rangle$ and the ancilla is prepared in $|A\rangle$, first the unitary map 
$$|\psi\rangle|A\rangle\to \frac{(1+\Pi)}{\sqrt{2}}|\psi\rangle|A\rangle+\frac{(1-\Pi)}{\sqrt{2}} |\psi\rangle|B\rangle\,$$ 
is applied; this map is the identity if the state possesses even parity, while if the state has no definite parity it becomes entangled with the ancilla. If in a subsequent measurement the ancilla is found in $|B\rangle$, conditional dynamics are performed that invert the parity of the system's state; in our simulations, a $\sigma_z$ operation is applied to one of the spins. Another option could be to reinitialize the spins to $|\up \ldots \up \rangle$. Denoting by $\mathcal P$ the corresponding map, the complete sequence of pulses we implement is ${\mathcal T}'=\mathcal P\mathcal T$. This protocol is efficient for {\it arbitrary} initial states and has constant depth, which makes it scalable \cite{Leibfried:unpublished}. It is conceptually an extension of methods for cooling the motion of ions based on measurements \cite{Eschner, Appasamy, Leibfried:12} (see also Ref.~\cite{Carvalho:14} for an application to QRE). Its realization requires that the coupling between spins and ancilla is homogeneous to a good degree.

\begin{figure}
\includegraphics[width=8.1cm]{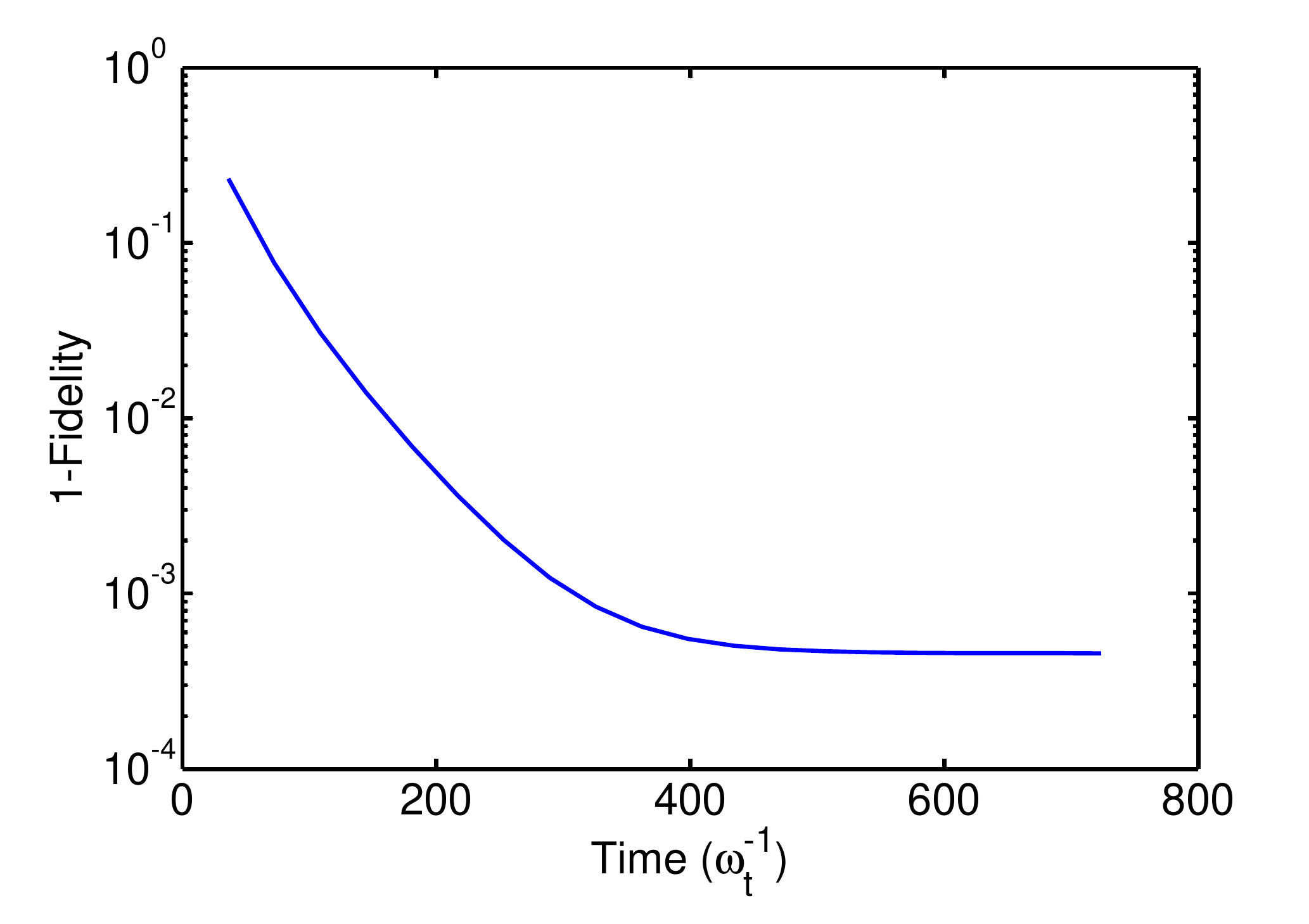}
\caption{Residual infidelity for preparing $N=4$ spins in the target state of Eq. \eqref{Target} for $\bar{n}=0$ and in absence of decoherence. The sequence $\mathcal T'$ contains pulses at the detunings $\Delta_1=-\omega_{\rm t}+2(J_1+J_2)$ (red arrows in Fig.~\ref{Fig:1}(b)), $\Delta_2=-\omega_{\rm t}+2J_1$ (blue), $\Delta_3=-\omega_t-2(J_1-J_2)$ (grey), $\Delta_4=-\Delta_1$ (green), and $\Delta_5=-\Delta_2$ (purple). The other parameters are $J_1=J_3=0.05\omega_{\rm t}$, $J_2=\sqrt{2}J_1$, $g_j=g=5\times 10^{-3}\omega_{\rm t}$, $\gamma t_{\gamma}=20$. The pulse durations are such that $J_1t_j>1$ ($J_1t_j\sim 10$), the pulse areas are optimized to minimize the loss rate $\Gamma_T$ \cite{Trapping}. The procedure is robust against parameter fluctuations: the infidelity doubles when the pulse areas change by about 30\%.}
\label{Fig:2}
\end{figure}

We now come to the specific features of an implementation based on Hamiltonian \eqref{H:int}, for the example of the target state of Eq.~\eqref{Target} with $N=2,4$ ions in a linear Paul trap. We first identify the frequencies, i.e. the detunings $\Delta_j$ of the pulses, for all transitions which couple the target state to any other state. For 2 ions, these are $\Delta_1=-\omega_{\rm t}+2J_1$ and $\Delta_2=-\Delta_1$, which pump $|\up\up\rangle|0\rangle\to |\psi_T\rangle|1\rangle$ and $|\down\down\rangle|0\rangle \to |\psi_T\rangle|1\rangle$, respectively; afterwards the ancilla is damped, making the transfer irreversible. The Rabi frequencies of the pulses are set to spectrally resolve the individual resonances; residual off-resonant coupling, which would depopulate the target state, is minimized by choosing pulse durations for which this coupling produces an integer number of Rabi oscillations. 
For this choice, we pump $N=2$ ions into the target state with fidelity $\mathcal F>1-10^{-6}$ under ideal conditions, i.e., for $\bar{n}=0$ and in absence of other sources of noise. For $N=4$ ions we identify a sequence of 5 pulses, shown in Fig.~\ref{Fig:1}(b) and detailed in the caption of Fig.~\ref{Fig:2}, which leads to an asymptotic fidelity $\mathcal F>0.9995$ under ideal conditions. The infidelity as a function of time is displayed in Fig.~\ref{Fig:2}.

\begin{figure}
\includegraphics[width=8.1cm]{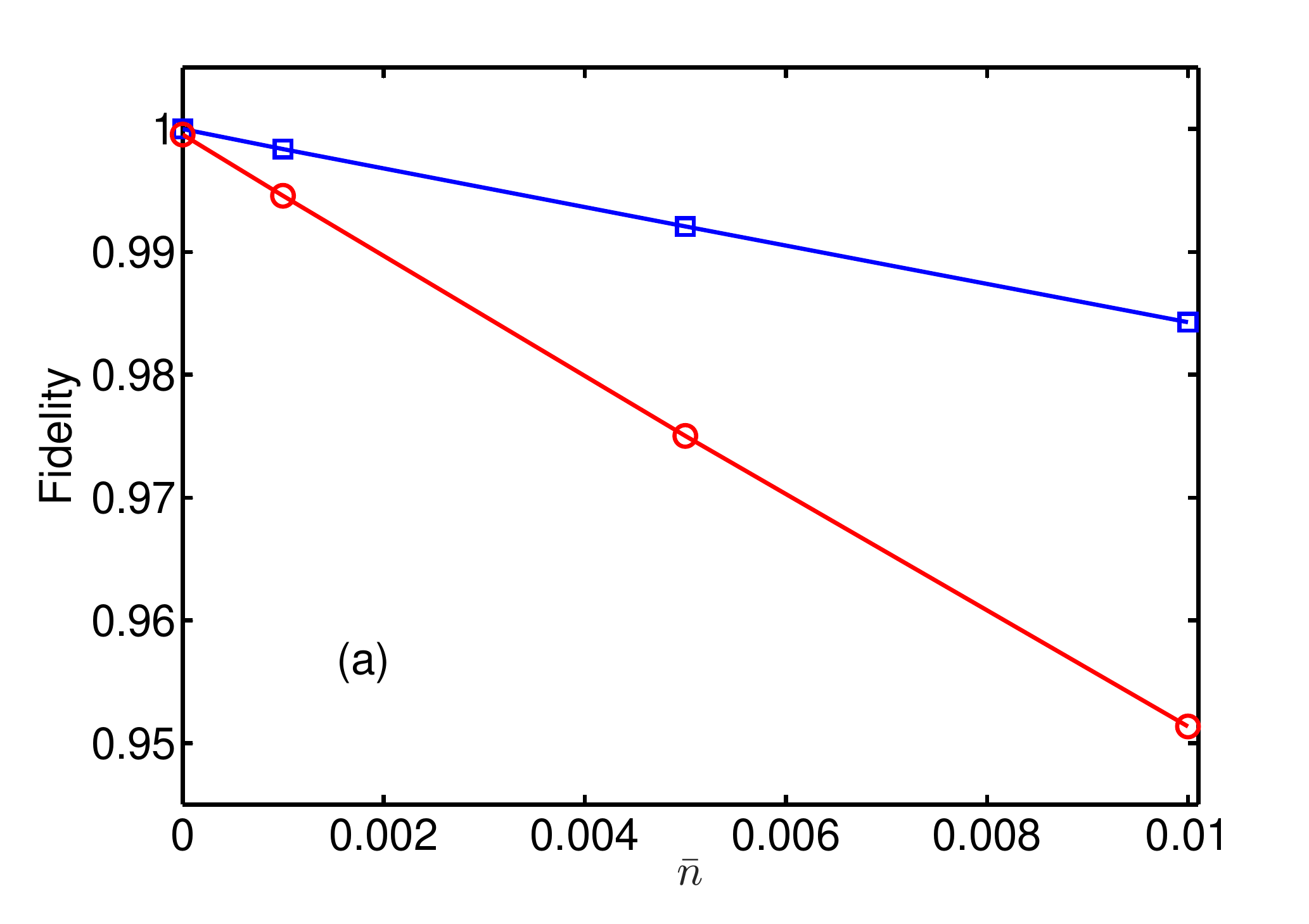}
\includegraphics[width=8.1cm]{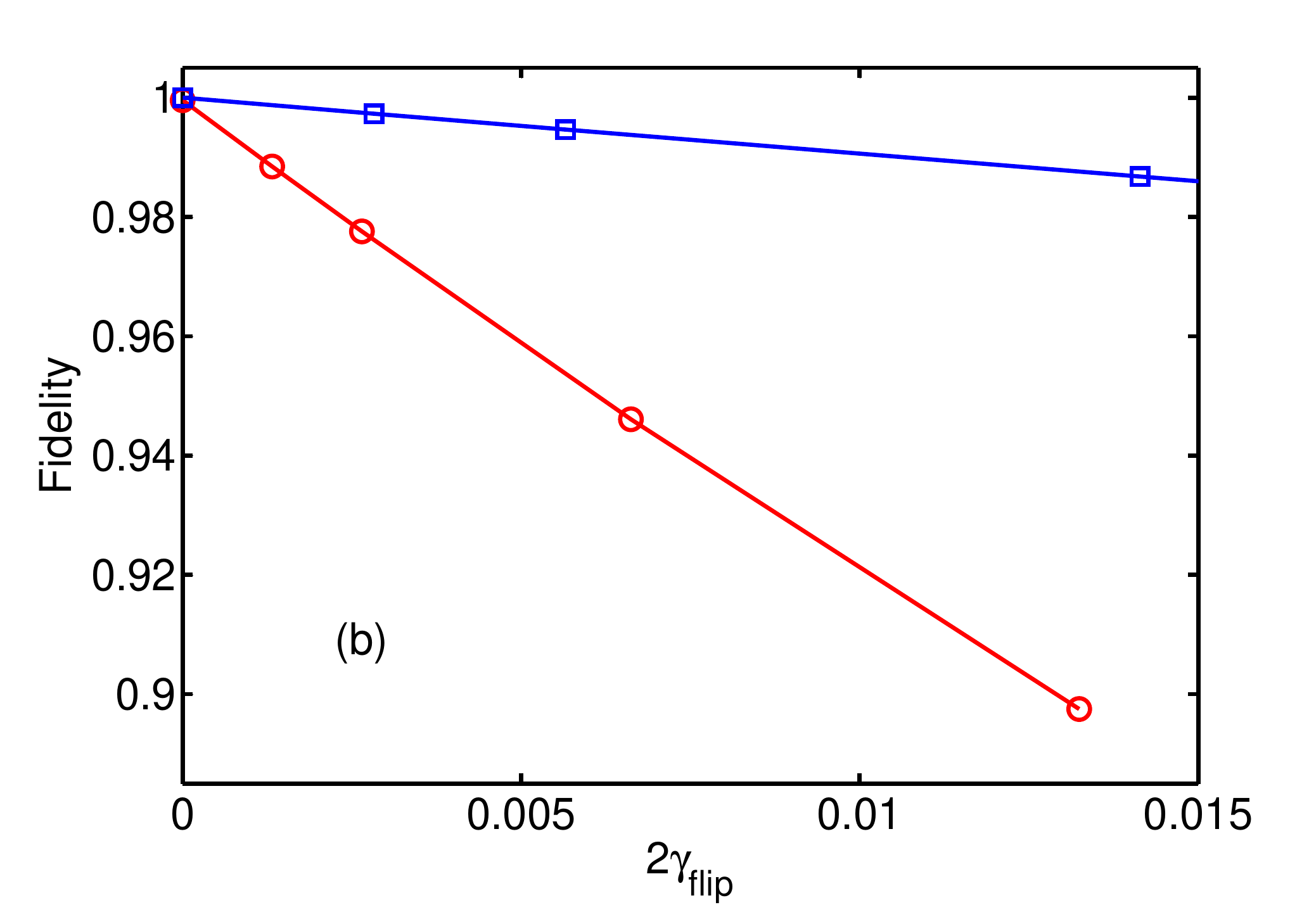}
\caption{Fidelity for preparing $N=2$ (blue squares) and $N=4$ (red circles) spins in the
target state as a function of (a) $\bar{n}$ (in absence of decoherence of the internal state) and (b)
the rate of incoherent processes $2\gamma_{\rm flip}$ in units of $g$ (for $\bar
n=0$ and assuming $\gamma_{\rm flip}=\gamma_{\rm deph}$). The fidelity
corresponds to the asymptotic value of a sequence of pulses with the map
${\mathcal T}'$. The fidelity is independent of the initial
state, provided the parity correction is included in the sequence. The
parameters are given in the caption of Fig.~\ref{Fig:2}.}
\label{Fig:3}
\end{figure}

We now analyse how the fidelity is affected by the ability to engineer the desired dissipation. We first consider varying the temperature of the reservoir and thus $\bar n$. Figure \ref{Fig:3}(a) displays the asymptotic fidelity for different values of $\bar n$ and for $N=2,4$ ions, and shows that the control on engineered dissipation becomes more stringent as the number of spins is increased. Figure \ref{Fig:3}(b) shows the fidelity in presence of noise and decoherence, which we consider here to be due to spin flips at rate $\gamma_{\rm flip}$  and dephasing with $\gamma_{\rm deph}$. The corresponding Liouvillians are $\gamma_{\rm flip}\sum_j\left(\mathcal D[\sigma_j]+\mathcal D[\sigma_j^\dagger]\right)$ and $\gamma_{\rm deph}\sum_j\mathcal D[\sigma_j^z]$, and are added to $\mathcal L_{\rm coh}(\Delta)$. The parity operation $\mathcal P$ counteracts the decoherence and keeps the fidelity above 0.9 for values of $\gamma_{\rm flip},\gamma_{\rm deph}$ for which in absence of $\mathcal P$ we observe a drop to 0.5. The curves show that the effect of noise becomes more detrimental as $N$ grows, which is expected as the protocol becomes slower because of the spectral crowding around the target state. This could be counteracted by increasing the energy splittings in the spectrum, here given by $\omega_t$ and $J_j$. It is important to note that the strength of the coupling, determining the speed of each pulse, scales differently with $N$ depending on the physical system. For instance, when the dissipative channel is the collective motion of an ion chain, the coupling between spins and motion decreases as $N$ grows, due to the increasing inertia of the crystal. If instead it is a cavity mode, the coupling can increase with $\sqrt{N}$ owing to superradiant emission.  

In conclusion, we have described a procedure for preparing a spin chain in an entangled state. The procotol uses engineered dissipation, which makes it robust against moderate fluctuations in the parameters, and includes a parity-error correction procedure, which makes it robust against detrimental noise and decoherence. The basic requirement is that the target state is spectrally resolved, which is achieved by constructing a suitable spin-spin interaction.  Efficient preparation of $N$-spin entangled states is warranted as long as the required spectral resolution is larger than the typical rate of noise and decoherence.  For arbitrary initial states, the protocol time scale is expected to increase exponentially with $N$, as it requires the capability to sweep over all state space. It can be notably reduced if the initial state is, e.g., a polarized chain such as $|\up,\up,\ldots,\up\rangle$, which is typically easy to produce with optical pumping. Its duration can be further shortened by optimizing the time duration of the coherent pulses, e.g., by using
time-dependent values of $\Delta$ and $g$ in Eq. \eqref{H:L} identified by means of optimal-control theory \cite{Koch}. 

G.M. and J.E. acknowledge the hospitality of the Ion Storage Group at NIST (Boulder) and support by the German DFG and BMBF (Q.Com). Y. L., D. L., and D. W. acknowledge support by IARPA, ONR and the NIST Quantum Information Program.


\begin{thebibliography}{widestlabel}

\bibitem{Horodecki}
R. Horodecki, P. Horodecki, M. Horodecki, and K, Horodecki,  
%Quantum entanglement,
Rev. Mod. Phys. {\bf 81}, 865 (2009).

\bibitem{Hillery}
M. Hillery, V. Buzek, and A. Berthiaume, %Quantum secret sharing, 
Phys. Rev. A {\bf 59}, 1829
(1999).

\bibitem{Josza}
R. Jozsa and N. Linden, %On the role of entanglement in quantum-computational speed-up, 
Proc. R. Soc. London A {\bf 459}, 2011
(2003).

\bibitem{Bruss}
D. Bruss and C.Macchiavello, %Multipartite entanglement in quantum algorithms,
Phys. Rev. A {\bf 83}, 052313 (2011).

\bibitem{Qsensor}
M. I. Kolobov, %The spatial behavior of nonclassical light, 
Rev. Mod. Phys. {\bf 71}, 1539 (1999); L. A. Lugiato, A. Gatti and E. Brambilla, %Quantum imaging, 
J. Opt. B %: Quantum Semiclassical Opt. 
{\bf 4}, S176 (2002).

\bibitem{Qmetrology}
V. Giovannetti, S. Lloyd, and L. Maccone, %Quantum Metrology,
Phys. Rev. Lett. {\bf 96}, 010401(2006); %Advances in quantum metrology, 
Nature Phot. {\bf 5}, 222 (2011).

\bibitem{Ladt}
T. D. Ladd, F. Jelezko, R. Laflamme, Y. Nakamura, C. Monroe, and  J. L. O'Brien,
%Quantum computers, 
Nature {\bf 464}, 45 (2010).


\bibitem{DallaTorre}
E. G. Dalla Torre, E. Demler, T. Giamarchi, and E. Altman, %Quantum critical states and phase transitions in the presence of non-equilibrium noise,
Nature Physics {\bf 6}, 806 (2010).

\bibitem{Poyatos}
J. F. Poyatos, J. I. Cirac, and P. Zoller, %Quantum Reservoir Engineering with Laser Cooled Trapped Ions, 
Phys. Rev. Lett.
{\bf 77}, 4728  (1996).

\bibitem{Diehl}
S. Diehl, A. Micheli, A. Kantian, B. Kraus, H. P. B\"uchler,
and P. Zoller, %Quantum states and phases in driven open quantum systems with cold atoms, 
Nature Phys. {\bf 4}, 878 (2008).

\bibitem{Verstraete}
F. Verstraete, M. M. Wolf, and J. I. Cirac, %Quantum computation and quantum-state engineering driven by dissipation, 
Nature Phys. {\bf 5}, 633 (2009),

\bibitem{Kraus}
B. Kraus, H. P. B\"uchler, S. Diehl, A. Kantian, A. Micheli, and P. Zoller,
%Preparation of entangled states by quantum Markov processes, 
Phys. Rev. A {\bf 78},
042307 (2008).

\bibitem{Kastler}
A. Kastler, %Quelques suggestions concernant la production
%optique et la d\'etection optique d'une in\'egalit\'e de population
%des niveaux de quantifigation spatiale des atomes, 
J. Phys. {\bf 11}, 255 (1950).

\bibitem{Davidovich}
L. Aolita, F. de Melo, and L. Davidovich, %Open-system dynamics of entanglement:a key issues review,
Rep. Prog. Phys. {\bf 78}, 042001 (2015)

\bibitem{Kim}
J. Cho, S. Bose, and M. S. Kim,  %Optical Pumping into Many-Body Entanglement,
Phys. Rev. Lett. {\bf 106}, 020504 (2011).

\bibitem{Viola}
F. Ticozzi and L. Viola, %Steady-State Entanglement by Engineered Quasi-Local Markovian Dissipation,
Quantum Information and Computation {\bf 14}, 0265 (2014).

\bibitem{Lidar}
D. A. Lidar, D. Bacon, and K. B. Whaley,  %Concatenating Decoherence-Free Subspaces with Quantum Error Correcting Codes,
Phys. Rev. Lett. {\bf 82}, 4556 (1999). 

\bibitem{Pielawa:07}
S. Pielawa, G. Morigi, D. Vitali, and L. Davidovich, 
%Generation of Einstein-Podolsky-Rosen-Entangled Radiation through an Atomic Reservoir,
Phys. Rev. Lett. {\bf 98}, 240401 (2007).

\bibitem{Davidovich:01}
A. R. R. Carvalho, P. Milman, R. L. de Matos Filho, and L. Davidovich,
%Decoherence, Pointer Engineering, and Quantum State Protection,
Phys. Rev. Lett. {\bf 86}, 4988 (2001).

\bibitem{PlenioHuelga}
M. B. Plenio and S. F. Huelga, %Entangled Light from White Noise, 
Phys. Rev. Lett. {\bf 88}, 197901 (2002).

\bibitem{Kastoryano} 
M. J. Kastoryano, F. Reiter, and A. S. S{\o}rensen, %Dissipative Preparation of Entanglement in Optical Cavities,
Phys. Rev. Lett. {\bf 106}, 090502 (2011). 

\bibitem{Molmer:13} 
D. D. Bhaktavatsala Rao and K. M{\o}lmer, 
%Dark Entangled Steady States of Interacting Rydberg Atoms,
Phys. Rev. Lett. {\bf 111}, 033606 (2013) .

\bibitem{CarrSaffman:13}
A. W. Carr and M. Saffman, 
%Preparation of Entangled and Antiferromagnetic States by Dissipative Rydberg Pumping,
Phys. Rev. Lett. {\bf 111}, 033607 (2013).

\bibitem{Cormick:13}
C. Cormick, A. Bermudez, S. F Huelga, and M. B Plenio, %Dissipative ground-state preparation of a spin chain by a structured environment , 
New J. Phys. {\bf 15}, 073027 (2013), 

\bibitem{Moelmer:14}
D. D. Bhaktavatsala Rao and K. M{\o}lmer,  
%Deterministic entanglement of Rydberg ensembles by engineered dissipation,
Phys. Rev. A {\bf 90}, 062319 (2014) 

\bibitem{Carvalho:14}
 C. D. B. Bentley, A. R. R. Carvalho, D. Kielpinski, and J. J. Hope,
%Detection-Enhanced Steady State Entanglement with Ions,
Phys. Rev. Lett. {\bf 113}, 040501 (2014).

\bibitem{Blatt}
J. T. Barreiro, M. M\"uller,	P. Schindler, D. Nigg, T. Monz, M. Chwalla, M.
Hennrich, C. F. Roos, P. Zoller, and R. Blatt, 
%An open-system quantum simulator with trapped ions, 
Nature {\bf 470}, 486 (2011). 

\bibitem{Lin}
Y. Lin, J. P. Gaebler, F. Reiter, T. R. Tan, R. Bowler, A. S. S{\o}rensen, D.
Leibfried, and D. J. Wineland, %Dissipative production of a maximally entangled steady state of two quantum bits, 
Nature {\bf  504}, 415 (2013).

\bibitem{Home}
D. Kienzler, H.-Y. Lo, B. Keitch, L. de Clercq, F. Leupold, F. Lindenfelser,
M. Marinelli, V. Negnevitsky, and J. P. Home, Science {\bf 347}, 53 (2015).

\bibitem{Krauter}
H. Krauter, C. A. Muschik, K. Jensen, W. Wasilewski, J. M. Petersen, J. I.
Cirac, and E. S. Polzik, %Entanglement Generated by Dissipation and Steady State
%Entanglement of Two Macroscopic Objects, 
Phys. Rev. Lett. {\bf 107}, 080503 (2011).

\bibitem{Devoret:2013}
S. Shankar, M. Hatridge, Z. Leghtas, K. M. Sliwa, A. Narla, U. Vool, S. M. Girvin, L. Frunzio, M. Mirrahimi, and M. H. Devoret,
%Autonomously stabilized entanglement between two superconducting quantum bits,
Nature {\bf 504}, 419 (2013).

\bibitem{Morigi07}
G. Morigi, P. W.H. Pinkse, M. Kowalewski, and R. de Vivie-Riedle, %Cavity Cooling of Internal Molecular Motion, 
Phys. Rev. Lett. {\bf 99}, 073001 (2007). 
%M. Kowalewski, G. Morigi, P. W. H. Pinkse, and R. de Vivie-Riedle, Appl. Phys. B 89, 459 (2007). 

\bibitem{Reiter:2015}
F. Reiter, D. Reeb, and A. S. S{\o}rensen, 
%Scalable dissipative preparation of many-body entanglement, 
preprint arXiv:1501.06611 (2015).


\bibitem{Dicke}
 R. H. Dicke, %Coherence in Spontaneous Radiation Processes,
Phys. Rev. {\bf 93}, 99 (1954).

\bibitem{Beige}
A. Beige, D. Braun, B. Tregenna, and P. L. Knight, %Quantum Computing Using Dissipation to Remain in a Decoherence-Free Subspace,
Phys. Rev. Lett. {\bf 85}, 1762 (2000).

\bibitem{Schaetz}
A. Bermudez, T. Schaetz, and M. B. Plenio, %Dissipation-Assisted Quantum Information Processing with Trapped Ions, 
Phys. Rev. Lett. {\bf 110}, 110502 (2013).

\bibitem{Sideband}
D. J. Wineland and Wayne M. Itano,  %Laser cooling of atoms,
Phys. Rev. A 20, 1521 (1979).

\bibitem{Malossi:2014}
N. Malossi, M. M. Valado, S. Scotto, P. Huillery, P. Pillet, D. Ciampini, E. Arimondo, and O. Morsch, %Full Counting Statistics and Phase Diagram of a Dissipative Rydberg Gas,
Phys. Rev. Lett. {\bf 113}, 023006 (2014).

\bibitem{Deng}
X.-L. Deng, D. Porras, and J. I. Cirac,
%Effective spin quantum phases in systems of trapped ions,
Phys. Rev. A {\bf 72}, 063407 (2005).

\bibitem{Footnote} State $|\psi_T\rangle$ is eigenstate of $H_0(\Delta)$ only for $N$ even. For $N$ odd, these conditions can be reached  by tailoring the transition frequency of the individual spins, for example by means of a spatial gradient of an external magnetic field as in F. Mintert and C. Wunderlich, %Ion-Trap Quantum Logic Using Long-Wavelength Radiation, 
Phys. Rev. Lett. {\bf 87}, 257904 (2001).

\bibitem{Gardiner}
C. W. Gardiner and P. Zoller, Quantum Noise (Springer-Verlag, Berlin, 2004).

\bibitem{LawEberly}
C. K. Law and J. H. Eberly, %Arbitrary Control of a Quantum Electromagnetic
Field, Phys. Rev. Lett. {\bf 76}, 1055 (1996).

\bibitem{Leibfried:04}
D. Leibfried, M. D. Barrett, T. Schaetz, J. Britton, J. Chiaverini, W.M. Itano,
J.D. Jost, C. Langer, and D.J. Wineland, %Toward Heisenberg-limited spectroscopy
%with multiparticle entangled states, 
Science {\bf 304}, 1476 (2004).

\bibitem{SorensenMolmer:99}
K. M{\o}lmer and A. S{\o}rensen, %Multiparticle Entanglement of Hot Trapped Ions,
Phys. Rev. Lett. {\bf 82}, 1835 (1999). 

\bibitem{Leibfried:unpublished}
D. Leibfried, unpublished.

\bibitem{Eschner}
J. Eschner, B. Appasamy, and P. E. Toschek, %Stochastic Cooling of a Trapped Ion by Null Detection of Its Fluorescence,
Phys. Rev. Lett. {\bf 74}, 2435 (1995). 

\bibitem{Appasamy}
B. Appasamy, Y. Stalgies, and P. Toschek, %Measurement-Induced Vibrational Dynamics of a Trapped Ion,
Phys. Rev. Lett. {\bf 80}, 2805 (1998).

\bibitem{Leibfried:12}
D. Leibfried, %Quantum state preparation and control of single molecular ions, 
New J. Phys. {\bf 14}, 023029 (2012). 

\bibitem{Trapping}
M. Weidinger, B. T. H. Varcoe, R. Heerlein, and H. Walther, 
%Trapping States inthe Micromaser,
Phys. Rev. Lett. {\bf 82}, 3795 (1999).

%\bibitem{Rojan}
%K. Rojan, D. M. Reich, I. Dotsenko, J.-M. Raimond, C. P. Koch, and G. Morigi,
%Arbitrary quantum-state preparation of a harmonic oscillator via optimal control,
%Phys. Rev. A {\bf 90}, 023824 (2014).

\bibitem{Koch}
D. M. Reich, C. P. Koch, %Cooling molecular vibrations with shaped laser pulses: Optimal control theory exploiting the timescale separation between
%coherent excitation and spontaneous emission, 
New J. Phys. {\bf 15}, 125028 (2013). 

%\bibitem{Fogarty}
%T. Fogarty, E. Kajari, B. G. Taketani, A. Wolf, T. Busch, and G. Morigi,
%%Entangling two defects via the surrounding crystal,
%Phys. Rev. A {\bf 87}, 050304 (2013).
%
%\bibitem{HuelgaRivas}
%S. F. Huelga, A. Rivas, and M. B. Plenio, 
%%Non-Markovianity-Assisted Steady State Entanglement,
%Phys. Rev. Lett. {\bf 108}, 160402 (2012).
% 
 
\end{thebibliography}
\end{document}